\documentclass[%
groupedaddress,
preprint,
showpacs,
amsmath,amssymb,
aps,
prb,
floats,
]{revtex4-1}

\usepackage{graphicx}
\usepackage{dcolumn}
\usepackage{bm}

\begin{document}


\title{Klein-tunneling-enhanced directional coupler for Dirac electron wave in graphene}%

\author{L. Zhao}
\email{zhaol@phys.tsinghua.edu.cn}
\affiliation{Department of Physics, Tsinghua University,
Beijing 100084, People's Republic of China }
\author{Wenhui Duan}%
\affiliation{Department of Physics, Tsinghua University,
Beijing 100084, People's Republic of China }

\date{\today}

\begin{abstract}
Using the coupled-mode theory in guided-wave optics and electronics, we explore a directional coupling structure composed of two parallel waveguides electrostatically induced by the split-gate technique in bulk graphene. Our results show that Klein tunneling can greatly enhance the coupling strength of the structure. By adjusting a gate voltage, the probability density of Dirac electron wave function initially in one waveguide can be completely transferred into the other waveguide within several hundred nanometers. Our findings could not only lead to functional coherent coupling devices for quantum-based electronic signal processing and on-chip device integration in graphene, but also shrink the size of the devices to facilitate the fabrication of graphene-based large-scale integrated logic circuits.
\end{abstract}

\maketitle


\section{Introduction}
Graphene, as a single layer of carbon atoms, has attracted massive attention since its successful isolation in 2004. \cite{slg0} The interaction between two-dimensional (2D) honeycomb carbon lattices and electrons in graphene generates massless Dirac fermions with a linear (relativistic) energy dispersion, which endows graphene with many fascinating electronic properties. \cite{gr1,gr2} For example, high carrier mobility with long coherence length ($\sim 1 \ \mu$m) has been reported in graphene. \cite{hm1,hm2,lcl} Another peculiar property is Klein tunneling which enables electrons in graphene to penetrate through high potential barriers with unity probability. \cite{k1} Many experimental observations associated with Klein tunneling have also been achieved. \cite{k2,k3}

Recently, the photon-like coherent transport behaviors of Dirac electrons in graphene, such as electron Veselago lensing, \cite{lens} electron beam supercollimation, \cite{coll} Goos-H\"{a}nchen-like shift, \cite{goos} and Fabry-P\'{e}rot interference, \cite{fp} have been studied extensively. Furthermore, to realize functional nanoelectronic devices for the applications of graphene-based integrated circuits, in analogy to the left-handed photonic waveguides in optics, \cite{tr} waveguide structures electrostatically induced by gate voltages in bulk graphene has been studied. \cite{wg1,wg2,wg3} In these waveguides, Klein tunneling can give rise to slow and even zero group velocity of the guided Dirac electrons, thereby leading to coherent memory devices for Dirac electrons. \cite{wg2} Additionally, it should be stressed that these electrostatically induced waveguide structures are totally different from the quasi-one-dimensional systems based on graphene nanoribbons. In the electrostatically induced waveguides in bulk graphene, the Dirac electrons are confined and guided by electrostatic potential barriers induced by gate voltages, which is also referred to as quantum wells in graphene. \cite{qw1} Whereas, in the graphene nanoribbons, the Dirac electrons are confined and guided by the edges of the ribbons, \cite{gr1,nr1} where atomic-scale tailoring processes remain a serious challenge to precisely control the transport of the Dirac electrons. \cite{nr2,nr3} Apparently, the electrostatically induced waveguides in bulk graphene avoid this challenge and thus could simplify the fabrication of nanoelectronic devices in graphene.

Nevertheless, the functionality of a single graphene waveguide is still limited and hard to satisfy increasing and diverse demand for electronic data processing. Therefore, complex quantum structures and devices in graphene should be explored, where a straightforward model is a directional coupler consisting of two parallel graphene waveguides. As a matter of fact, directional coupling structures have been well established both theoretically and experimentally using coupled-mode theory in guided-wave optics for the applications in integrated photonic circuits, including optical signal division, switching, multiplexing, and demultiplexing. \cite{yariv} As an electronic counterpart, quantum-field-effect directional coupler for coherent electron wave have also been intensively investigated using the split-gate technique in conventional 2D electron gas in AlGaAs/GaAs heterostructures. \cite{dc1,dc2,dc3} However, considering the exceptional properties of Dirac electrons in graphene, it is still intriguing to examine the coherent transport behaviors of Dirac electrons in coupled dual waveguide structures. Moreover, significant progress has been made in developing graphene-related nanotechnologies. The recent synthesis of large-area graphene sheets with high quality \cite{la1,la2} and the successful fabrication of nanoscale local electrostatic gates for graphene {\it p-n-p} junctions \cite{k3} make it experimentally possible to construct dual waveguide configurations with top and back gates in bulk graphene.

\begin{figure}[ht]
   \centerline{ \includegraphics[clip,width=0.8\linewidth]{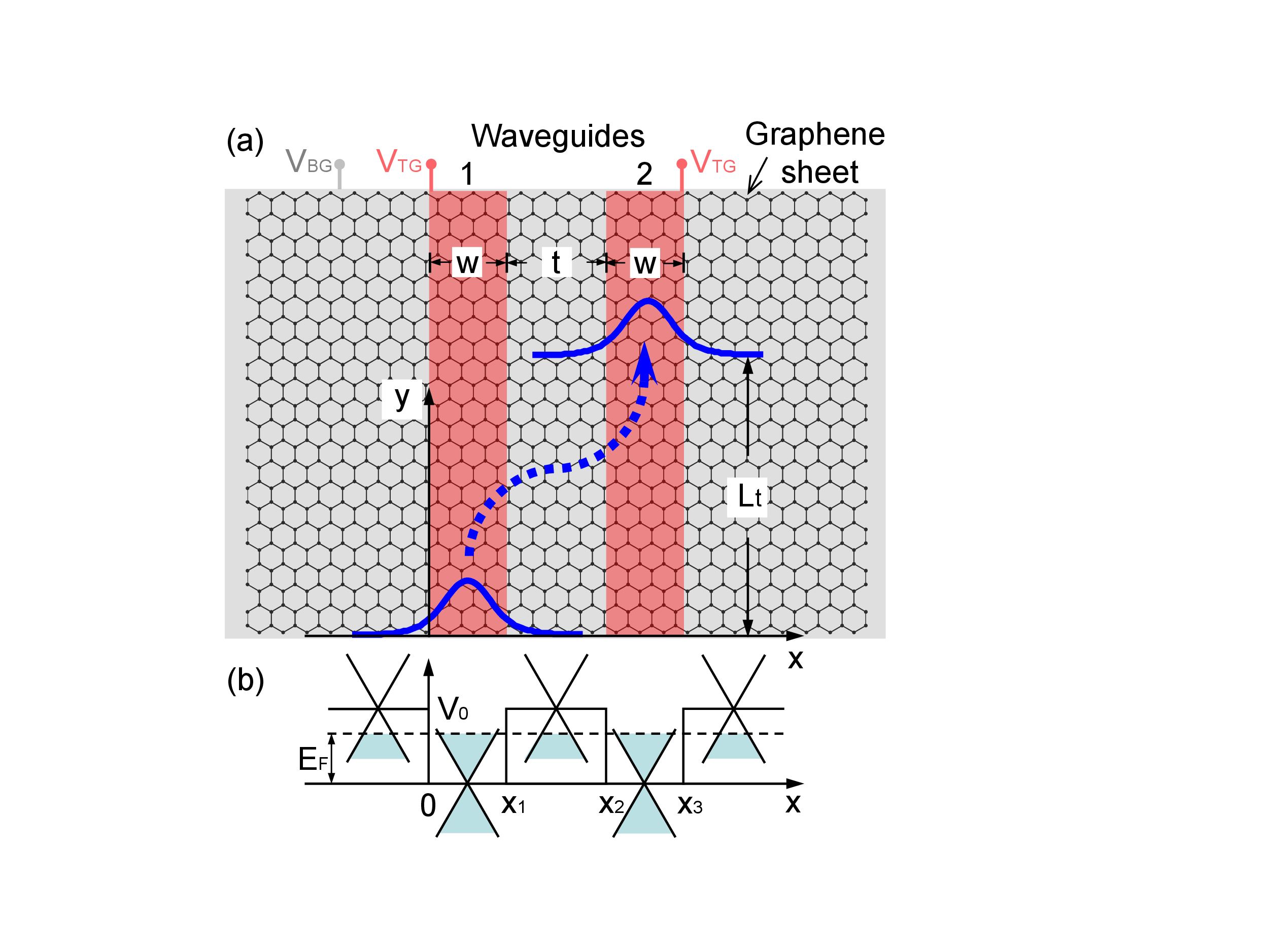}}
         \caption{(Color online) (a) Schematic diagram of a directional coupler consisting of two parallel waveguides electrostatically induced by the top ($V_{TG}$) and back ($V_{BG}$) gates in a large-area graphene sheet. (b) Potential diagram and energy spectrum of the coupler, where we only show the {\it n-p} interfaces for Klein tunneling. For simplicity, we consider a symmetric configuration with two identical waveguides having the same width $w$, potential barrier height $V_{0}$, and Fermi energy $E_{F}$. And, the separation distance between the two waveguides is $t$. The shortest length for a complete wave function transfer is $L_{t}$. }
\label{fig:coupler}
\end{figure}

In this paper, we utilize the coupled-mode theory in guided-wave optics \cite{yariv} and electronics \cite{dc1,dc2,dc3} to investigate the coherent transport of Dirac electrons in a directional coupler electrostatically induced by the split-gate technique in a bulk graphene sheet [Fig. \ref{fig:coupler}(a)]. Our results indicate that Klein tunneling can dramatically enhance the coupling strength of the system. Due to this enhancement, the probability density of Dirac electron wave function initially injected into one waveguide can be completely transferred into the other waveguide within a propagation length of several hundred nanometers by tuning a gate voltage. It is notable that this length is smaller than the coherence length ($\sim 1 \ \mu$m) observed in graphene. \cite{lcl} In practice, this enhancement also means that wave function transfer can be accomplished in a compact geometry, thereby shrinking the size of graphene-based directional coupling devices. As well as understanding a fundamental interacting phenomenon enhanced by Klein tunneling between two Dirac-type quantum objects, our study could provide a functional quantum coupling device for nanoelectronic applications in grapnene, such as Dirac electron switching, splitting, multiplexing and demutiplexing.

\section{Theoretical model}

In our scheme, for simplicity, we assume that the two waveguides 1 and 2 have identical square potential profile [Fig. \ref{fig:coupler}(b)]. Following the idea of the coupled-mode theory, \cite{yariv,dc1,dc2} we first consider the two waveguides separately. For the {\it isolated} waveguide 1, its potential can be give by
\begin{equation}\label{pot1}
V_{1}(x)=\left\{\begin{array}{ll}
0,&\ 0\le x \le x_{1}, \\ V_{0}, & \ \mathrm{otherwise}.
         \end{array}\right.
\end{equation}
The guided Dirac electron wave function is governed by the 2D Dirac equation, \cite{gr1,gr2,qw1,wg1,wg2}
\begin{equation} \label{dirac1}
[-i\hbar v_{F} (\boldsymbol{\sigma} \cdot \boldsymbol{\nabla})+V_{1}(x)] |\psi_{1} \rangle= E_{1} |\psi_{1}\rangle,
\end{equation}
where $v_{F} \approx 10^{6}$ m/s is the Fermi velocity, $\boldsymbol{\sigma}=(\sigma_{x},\sigma_{y})$ are the Pauli matrices, $\boldsymbol{\nabla} = (\partial/\partial x,\partial/\partial y)$, the eigenenergy is $E_{1}=E_{F}$, and $E_{F}$ is the Fermi energy of waveguides 1. Due to the translational invariance along the $y$ direction, $| \psi_{1} \rangle$ can be assumed as $| \psi_{1}(x,y) \rangle= | \phi_{1}(x) \rangle e^{i k_{y} y}$, where $k_{y}$ is the momentum in the $y$ direction. Moreover, because graphene honeycomb lattice contains two sublattices A and B, $|\phi_{1}(x) \rangle$ can be expressed by a two-component spinor, i.e., $|\phi_{1}(x) \rangle=[\phi_{1}^{A}(x),\phi_{1}^{B}(x)]^{T}$, where $\phi_{1}^{A}(x)$ and $\phi_{1}^{B}(x)$ represent the smooth enveloping functions in each sublattice. Therefore, Eq. (\ref{dirac1}) can be expressed by \cite{qw1,wg1,wg2}
\begin{subequations} \label{diracphi}
\begin{eqnarray}
\frac{d \phi_{1}^{A}(x)}{d x}-k_{y}\phi_{1}^{A}(x) &=& i\frac{E_{1}-V_{1}(x)}{\hbar v_{F}}\phi_{1}^{B}(x), \label{diracphi1} \\
\frac{d \phi_{1}^{B}(x)}{d x}+k_{y}\phi_{1}^{B}(x) &=& i\frac{E_{1}-V_{1}(x)}{\hbar v_{F}}\phi_{1}^{A}(x). \label{diracphi2}
\end{eqnarray}
\end{subequations}
For the isolated waveguide 1, the wave functions can be specifically expressed as follows. For $x < 0$, we have \cite{qw1,wg1,wg2}
\begin{eqnarray}\label{wf11}
|\phi_{1}\rangle &=& \binom{\phi_{1}^{A}(x)}{\phi_{1}^{B}(x)}=
a \binom{1}{ i(k_{y}-\alpha)/k_{2}} e^{\alpha x},
\end{eqnarray}
where $k_{2}=(E_{F}-V_{0})/\hbar v_{F}$ and $\alpha^{2}=k_{y}^{2}-k_{2}^{2}$.
For $0 \le x \le x_{1}$, we have \cite{qw1,wg1,wg2}
\begin{eqnarray}\label{wf12}
|\phi_{1}\rangle =\binom{b \cos(k_{x}x)+c \sin(k_{x}x)}
{i[(b+c)k_{x} \sin (k_{x} x) + (b-c)k_{y} \cos (k_{x} x)]/k_{1}}, \nonumber \\
\end{eqnarray}
where $k_{1}=E_{F}/\hbar v_{F}$, $k_{x}^{2}=k_{1}^{2}-k_{y}^{2}$, $x_{1}=w$, and $w$ is the width of an individual waveguide.
For $x > x_{1}$, we have \cite{qw1,wg1,wg2}
\begin{eqnarray}\label{wf13}
|\phi_{1}\rangle &=& d \binom{1}{i(\alpha+k_{y})/k_{2}}  e^{-\alpha (x-x_{1})}.
\end{eqnarray}
The parameters $a$, $b$,$c$,and $d$ in the above equations are the normalization coefficients. Moreover, to obtain the guided modes in the waveguide, the total internal reflection at waveguide interfaces is required, which results in the condition of $0 < V_{0} < 2 E_{F}$. For $0 < V_{0} < E_{F}$, we have {\it n-n$^{\prime}$} interfaces with intraband tunneling. For $E_{F} < V_{0} < 2E_{F}$, we have {\it n-p} interfaces with Klein tunneling. Additionally, using the continuity conditions of the wave functions at $x=0$ and $x=x_{1}=w$, we can obtain \cite{qw1,wg1,wg2}
\begin{eqnarray}\label{boundary}
\tan(k_{x}w)=\frac{k_{x} \sqrt {k_{1}^{2}-k_{x}^{2}-k_{2}^{2}}}{k_{1}k_{2}-(k_{1}^{2}-k_{x}^{2})},
\end{eqnarray}
which is usually used to determine $k_{x}$ of the discrete guided eigenmodes in the waveguide.

Similarly, for the {\it isolated} waveguide 2, its potential is
\begin{equation}\label{pot2}
V_{2}(x)=\left\{\begin{array}{ll}
0,&\ x_{2}\le x \le x_{3}, \\ V_{0}, & \ \mathrm{otherwise}.
         \end{array}\right.
\end{equation}
The guided Dirac electron wave function is governed by the 2D Dirac equation, \cite{gr1,gr2,qw1,wg1,wg2}
\begin{equation} \label{dirac2}
[-i\hbar v_{F} (\boldsymbol{\sigma} \cdot \boldsymbol{\nabla})+V_{2}(x)] |\psi_{2} \rangle= E_{2} |\psi_{2}\rangle,
\end{equation}
where the eigenenergy is $E_{2} = E_{1}=E_{F}$ for the identical waveguides. The guided Dirac electron wave function can be given by $|\psi_{2}(x,y) \rangle = |\phi_{2}(x) \rangle e^{i k_{y} y}=[\phi_{2}^{A}(x),\phi_{2}^{B}(x)]^{T}e^{i k_{y} y}$. For the isolated waveguide 2, the wave functions can be specifically expressed as follows.
For $x < x_{2}$, we have \cite{qw1,wg1,wg2}
\begin{eqnarray}\label{wf21}
|\phi_{2}\rangle &=& \binom{\phi_{2}^{A}(x)}{\phi_{2}^{B}(x)}=
a \binom{1}{ i(k_{y}-\alpha)/k_{2}} e^{\alpha x^{\prime}},
\end{eqnarray}
where $x^{\prime}=x-x_{2}$, $x_{2}=w+t$, and $t$ is the separation distance between the two waveguides. For $x_{2} \le x \le x_{3}$, we have \cite{qw1,wg1,wg2}
\begin{eqnarray}\label{wf22}
|\phi_{2}\rangle =\binom{b \cos(k_{x}x^{\prime})+c \sin(k_{x}x^{\prime})}
{i[(b+c)k_{x} \sin (k_{x}x^{\prime})+ (b-c)k_{y} \cos (k_{x} x^{\prime})]/k_{1}}. \nonumber \\
\end{eqnarray}
For $x > x_{3}$, we have \cite{qw1,wg1,wg2}
\begin{eqnarray}\label{wf23}
|\phi_{2}\rangle &=& d \binom{1}{i(\alpha+k_{y})/k_{2}}  e^{-\alpha x^{\prime \prime}},
\end{eqnarray}
where $x^{\prime \prime}=x-x_{3}$ and $x_{3}=2w+t$.

When the two waveguides are very close to each other, the wave functions will overlap in the $x$ direction due to quantum tunneling, which can affect the propagation of the wave functions in the $y$ direction. Therefore, based on coupled-mode theory, \cite{yariv,dc1,dc2,dc3} we assume that the wave function in the dual waveguides is a linear combination of the wave functions in each isolated waveguide, i.e.,
\begin{equation}\label{twf}
| \psi \rangle=p(y) | \psi_{1} \rangle+q(y) | \psi_{2} \rangle,
\end{equation}
where $p(y)$ and $q(y)$ characterize the propagation of $\psi$ in the $y$ direction. The wave function should satisfy the 2D Dirac equation for the dual waveguide coupler. That is \cite{gr1,gr2,dc1,dc2,dc3}
\begin{eqnarray} \label{dirac12}
[-i\hbar v_{F} (\boldsymbol{\sigma} \cdot \boldsymbol{\nabla})+V(x)] | \psi \rangle \;=\; E_{F} |\psi \rangle,
\end{eqnarray}
where $V(x)=V_{1}(x)+V_{2}(x)-V_{0}$. Using Eqs. (\ref{dirac1}) and (\ref{dirac2}), Eq. (\ref{dirac12}) can be transformed into the form of
\begin{eqnarray} \label{dirac3}
\frac{d p}{d y} |\phi_{1} \rangle +\frac{d q}{d y} | \phi_{2} \rangle =
p \frac{V_{2}^{\prime}}{i\hbar v_{F}} \sigma_{y} |\phi_{1}\rangle +q\frac{V_{1}^{\prime}}{i\hbar v_{F}} \sigma_{y} |\phi_{2}\rangle,
\end{eqnarray}
where $V_{2}^{\prime}=V_{2}-V_{0}$ and $V_{1}^{\prime}=V_{1}-V_{0}$. Taking the inner product with $\langle \phi_{1}|$, Eq. (\ref{dirac3}) can be rewritten as
\begin{eqnarray} \label{dirac31}
& & \frac{d p}{d y} \langle \phi_{1}|\phi_{1} \rangle +\frac{d q}{d y} \langle \phi_{1}| \phi_{2} \rangle = \nonumber    \\
& & \frac{p}{i\hbar v_{F}} \langle \phi_{1} | V_{2}^{\prime} \sigma_{y} |\phi_{1}\rangle +\frac{q}{i\hbar v_{F}}\langle \phi_{1}|V_{1}^{\prime} \sigma_{y} |\phi_{2}\rangle.
\end{eqnarray}
Likewise, taking the inner product with $\langle \phi_{2}|$, Eq. (\ref{dirac3}) can be rewritten as
\begin{eqnarray} \label{dirac32}
& & \frac{d p}{d y} \langle \phi_{2}|\phi_{1} \rangle +\frac{d q}{d y} \langle \phi_{2}| \phi_{2} \rangle = \nonumber    \\
& & \frac{p}{i\hbar v_{F}} \langle \phi_{2} | V_{2}^{\prime} \sigma_{y} |\phi_{1}\rangle +\frac{q}{i\hbar v_{F}}\langle \phi_{2}|V_{1}^{\prime} \sigma_{y} |\phi_{2}\rangle.
\end{eqnarray}
For the normalized wave functions $| \phi_{1} \rangle$ and $| \phi_{2} \rangle$ under weak-coupling condition between the waveguides, the overlap and tunneling of the wave functions $|\phi_{1} \rangle$ and $|\phi_{2} \rangle$ are weak. \cite{yariv,dc1,dc2} Therefore, one can define the parameters $R_{1}$, $R_{2}$, $R_{3}$, and $R_{4}$ as follows to describe the weak-coupling condition
\begin{subequations} \label{weak}
\begin{eqnarray}
R_{1}&=&\frac{|\langle \phi_{1}|\phi_{2} \rangle|}{|\langle \phi_{1}|\phi_{1} \rangle|} \ll 1, \label{weak1} \\
R_{2}&=&\frac{|\langle \phi_{1} | V_{2}^{\prime} \sigma_{y} |\phi_{1}\rangle|}
{|\langle \phi_{1} | V_{1}^{\prime} \sigma_{y} |\phi_{2}\rangle|} \ll 1,\label{weak2} \\
R_{3}&=&\frac{|\langle \phi_{2}|\phi_{1} \rangle|}{|\langle \phi_{2}|\phi_{2} \rangle|} \ll 1, \label{weak3} \\
R_{4}&=&\frac{|\langle \phi_{2} | V_{1}^{\prime} \sigma_{y} |\phi_{2}\rangle|}
{|\langle \phi_{2} | V_{2}^{\prime} \sigma_{y} |\phi_{1}\rangle|} \ll 1, \label{weak4}
\end{eqnarray}
\end{subequations}
where we have $|\langle \phi_{1}|\phi_{1} \rangle|=|\langle \phi_{2}|\phi_{2} \rangle|=1$ for the normalized wave functions $|\phi_{1} \rangle$ and $|\phi_{2} \rangle$. Due to the potential profiles of our configuration, Eqs. (\ref{weak2}) and (\ref{weak4}) can be further simplified to
\begin{subequations} \label{weak24}
\begin{eqnarray}
R_{2}&=&\frac{|\langle \phi_{1} | \sigma_{y} |\phi_{1}\rangle|_{x_{2} \le x \le x_{3}}}{|\langle \phi_{1} | \sigma_{y} |\phi_{2}\rangle|_{0 \le x \le x_{1}}} \ll 1, \label{sweak2} \\
R_{4}&=&\frac{|\langle \phi_{2} | \sigma_{y} |\phi_{2}\rangle|_{0 \le x \le x_{1}}} {|\langle \phi_{2} |\sigma_{y} |\phi_{1}\rangle|_{x_{2} \le x \le x_{3}}} \ll 1. \label{sweak4}
\end{eqnarray}
\end{subequations}
The validity of these conditions will be numerically verified latter in Section III. Thus, based on the weak-coupling approximation of Eqs. (\ref{weak}), Eqs. (\ref{dirac31}) and (\ref{dirac32}) become the coupled equations
\begin{subequations} \label{couple}
\begin{eqnarray}
\frac{d p}{d y} &=& \kappa_{12} q, \\
\frac{d q}{d y} &=& \kappa_{21} p,
\end{eqnarray}
\end{subequations}
where
\begin{subequations} \label{kappa}
\begin{eqnarray}
\kappa_{12} &=& \frac{\langle \phi_{1} | V_{1}^{\prime} \sigma_{y} | \phi_{2} \rangle}{i \hbar v_{F}}, \\
\kappa_{21} &=&  \frac{\langle \phi_{2} | V_{2}^{\prime} \sigma_{y} | \phi_{1} \rangle} {i \hbar v_{F}}.
\end{eqnarray}
\end{subequations}
Here the parameters $|\kappa_{12}|$ and $|\kappa_{21}|$ are the coupling coefficients between the two waveguides, which characterize the coupling strength of the directional coupler. In our symmetric configuration, we have $|\kappa|=|\kappa_{12}|=|\kappa_{21}|$. For the initial condition of $p(0)=1$ and $q(0)=0$ (i.e., the Dirac electron is initial injected into waveguide 1), the solutions of Eq. (\ref{couple}) are expressed by
\begin{subequations} \label{sol}
\begin{eqnarray}
|p(y)|^{2}&=&\cos^{2}(|\kappa|y), \\
|q(y)|^{2}&=&\sin^{2}(|\kappa|y),
\end{eqnarray}
\end{subequations}
which indicate a sinusoidal oscillation of the probability density of the guided Dirac electron wave between the two waveguides with the propagation in the $y$ direction. The shortest length to achieve a complete wave function transfer is given by
\begin{eqnarray} \label{lt}
L_{t}= \frac{\pi}{2 |\kappa|}.
\end{eqnarray}

\section{Numerical Results and Discussion}

\begin{figure}[ht]
\centerline{\includegraphics[clip,width=0.9\linewidth]{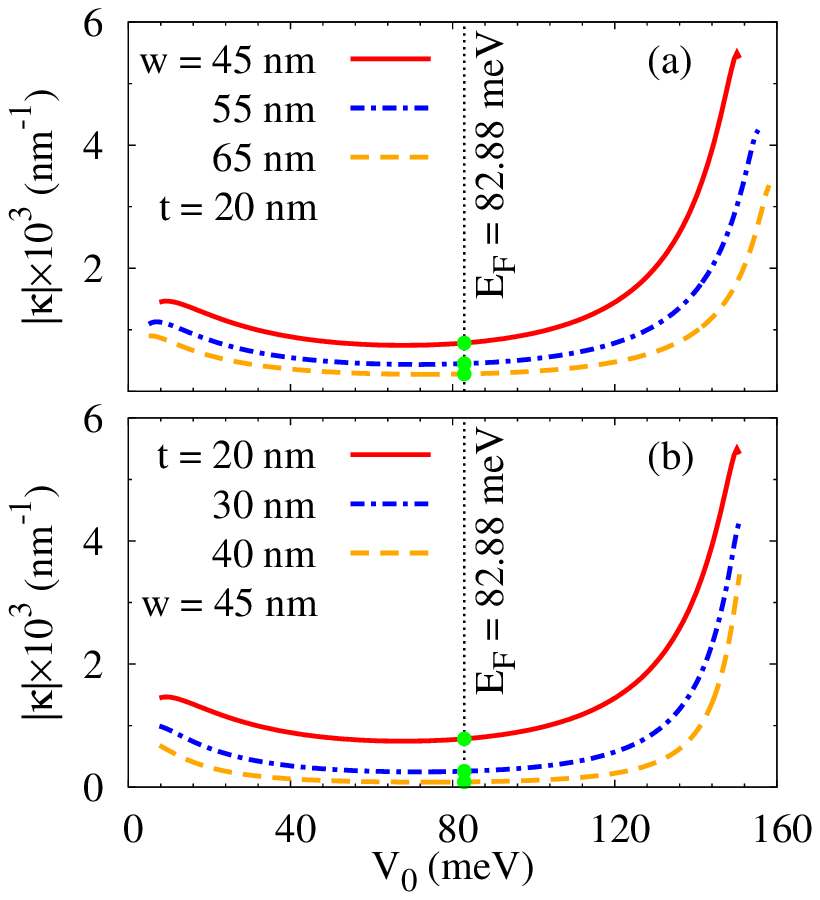}}
\caption{(Color online) (a) Coupling coefficients $|\kappa|$ for different waveguide widths of $w=45$ nm (solid curve), $55$ nm (dash-dotted curve), and $65$ nm (dashed curve), where the Fermi energy $E_{F}=82.88$ meV (dotted line) and the separation distance $t=20$ nm are fixed. (b) Coupling coefficients $|\kappa|$ for different separation distances of $t=20$ nm (solid curve), $30$ nm (dash-dotted curve), and $40$ nm (dashed curve), where the Fermi energy $E_{F}=82.88$ meV (dotted line) and the waveguide width $w=45$ nm are fixed. For $0<V_{0}<E_{F}$ on the left side, we have {\it n-n$^{\prime}$} interfaces with intraband tunnelling. For $E_{F}<V_{0}<2E_{F}$ on the right side, we have {\it n-p} interfaces with Klein tunneling. For all the curves, it is clearly seen that the coupling coefficients can be dramatically enhanced by Klein tunneling in comparison to intraband tunneling.}
\label{fig:cc}
\end{figure}

To evaluate the coupling strength between the two waveguides, we numerically investigate their coupling coefficient $|\kappa|$ using the derivations in Section II. In the numerical calculations, for simplicity, we fix the Fermi energy of the system at $E_{F}=82.88$ meV and assume that only the fundamental modes of the probability density with the minimum $k_{x}$ given by Eq. (\ref{boundary}) are excited in the waveguides. Comparing different curves in Fig. \ref{fig:cc}, at the same potential barrier height $V_{0}$, one can see that narrower waveguide widths $w$ or smaller separation distances $t$ can lead to larger coupling coefficients $\kappa$. These results originates from the fact that narrower waveguide widths $w$ or smaller separation distances $t$ can cause more wave functions to tunnel into the other waveguide, thereby increasing the coupling between the waveguides. However, for each individual curve, the Klein tunneling (i.e., $E_{F}<V_{0}<2E_{F}$) can lead to much larger coupling coefficients than the intraband tunneling (i.e., $0<V_{0}<E_{F}$). This gives an {\it unexpected} result (i.e., higher potential barriers can induce stronger coupling) in contrast to that in conventional AlGaAs/GaAs electronic systems. \cite{dc1,dc2,dc3}

\begin{figure*}[ht]
\centerline{\includegraphics[clip,width=0.9\linewidth]{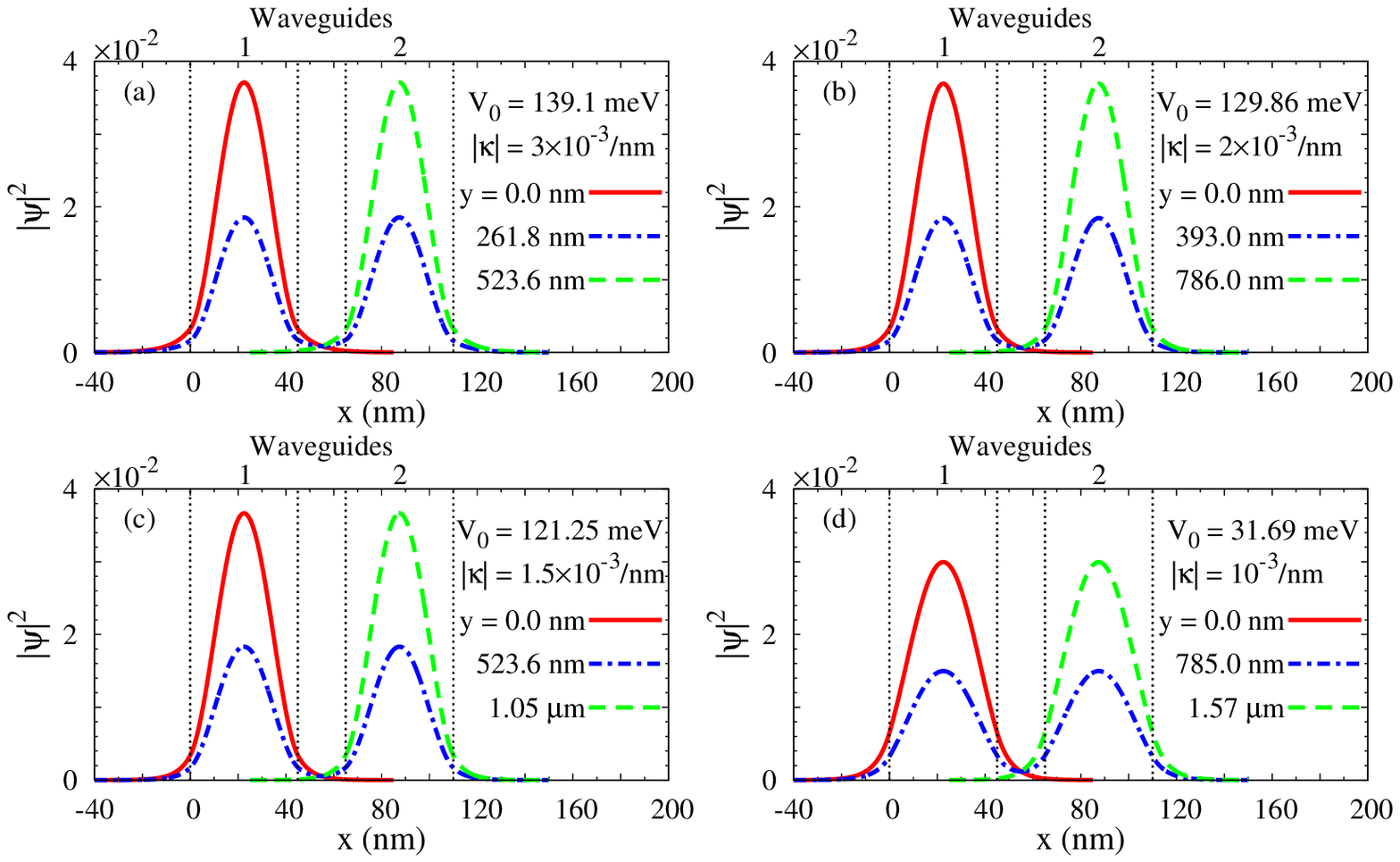}}
\caption{(Color online) Spatial transverse distributions of the normalized probability densities at different propagation lengths along the $y$ direction. Note that, for simplicity, we assume that only the fundamental modes of the probability density are excited in the waveguides. Here, we consider the directional couplers with the same waveguide width of $w=45$ nm, the same separation distance of $t=20$ nm and the same Fermi energy $E_{F}=82.88$ meV, but different potential barrier heights of (a) $V_{0}=139.1$ meV, (b) $V_{0}=129.86$ meV, (c) $V_{0}=121.25$ meV, and (d) $V_{0}=31.69$ meV, where the dotted lines indicate the positions of the two waveguides. In (a), we can obtain the coupling coefficient $|\kappa|=3 \times 10^{-3}$ nm$^{-1}$. The Dirac electron is initially injected into waveguide 1 (i.e., $y=0.0$ nm, see solid curve). After propagating $y=261.8$ nm, half of the probability density will be transferred into waveguide 2 (dash-dotted curve) due to the Klein tunneling between the two waveguides, where probability density splitting can be achieved. After propagating $y=523.6$ nm, the probability density will be completely transferred into waveguide 2 (dashed curve). In (b), (c), and (d), we can obtain the coupling coefficients $|\kappa|=2 \times 10^{-3}$ nm$^{-1}$, $|\kappa|=1.5 \times 10^{-3}$ nm$^{-1}$, and $|\kappa|= 10^{-3}$ nm$^{-1}$, respectively. Similar transport phenomena will occur at different propagation lengths. However, it is notable that£¬ compared with the interband tunneling (i.e., $0<V_{0}<E_{F}$) in (d), the Klein tunneling (i.e., $E_{F}<V_{0}<2E_{F}$) in (a), (b), and (c) can dramatically enhance the coupling strength between the two waveguides, thereby shrinking the propagation length $y$ for probability density transfer.
}
\label{fig:pd}
\end{figure*}

In practice, when a directional coupler is fabricated in graphene, the waveguide width $w$ and the separation distance $t$ are unchangeable. But, by tuning a gate voltage, the potential barrier height $V_{0}$ can be adjusted to control the transfer of Dirac electron wave between the two waveguides. For example, we here consider the directional coupler with $w=45$ nm, $t=20$ nm and $E_{F}=82.88$ meV [see the solid curve in Fig. \ref{fig:cc}(a)]. In this coupler, we use four different potential barrier heights $V_{0}$ to perform our numerical evaluations.

(i) At $V_{0}=139.1$ meV (for Klein tunneling at the waveguide interfaces), we can find $|\kappa|=3 \times 10^{-3}$ nm$^{-1}$ and thus the shortest length for a complete probability density transfer is $L_{t}=523.6$ nm given by Eq. (\ref{lt}). Note that $L_{t}$ is smaller than the observed coherence length ($\sim 1$ $\mu$m) of Dirac electrons in graphene. \cite{lcl} In Fig. \ref{fig:pd}(a), one can clearly see the spatial evolution of the probability density of the guided Dirac electron along the propagation (i.e., $y$) direction. The wave function is initially injected into waveguide 1 (i.e., the propagation length $y=0.0$ nm). At the propagation length $y=L_{t}/2=261.8$ nm, the probability density will be equally split into the two waveguides. At the propagation length $y=523.6$ nm, the probability density will be completely transferred into waveguide 2.

(ii) At $V_{0}=129.86$ meV (for Klein tunneling), we can find $|\kappa|=2 \times 10^{-3}$ nm$^{-1}$ and thus the shortest length for a complete probability density transfer is $L_{t}=786.0$ nm.

(iii) At $V_{0}=121.25$ meV (for Klein tunneling), we can find $|\kappa|=1.5 \times 10^{-3}$ nm$^{-1}$ and thus the shortest length is $L_{t}=1.05$ $\mu$m.

(iv) For comparison, at $V_{0}=31.69$ meV (for intraband tunneling), we can find $|\kappa|=10^{-3}$ nm$^{-1}$ and thus the shortest length is $L_{t}=1.57$ $\mu$m which is much longer than that in case (i).

In Figs. \ref{fig:pd}(b), \ref{fig:pd}(c) and \ref{fig:pd}(d), one can also see the spatial evolution of the probability density of the guided Dirac electron along the propagation (i.e., $y$) direction for (ii), (iii), and (iv), respectively. Therefore, Klein tunneling can dramatically enhance the coupling strength between the two waveguides. By tuning a gate voltage, one can fully control the spatial evolution of the probability density distribution between the two waveguides.

Based on the above results, a compact voltage-controlled coherent device enhanced by Klein-tunneling effect can be proposed. For example, we can assume that a directional coupler has a total length of $523.6$ nm in the $y$ direction and its transverse geometric parameters are still $w=45$ nm and $t=20$ nm. If the Dirac electron is initially injected into waveguide 1 at $V_{0}=139.1$ meV, it will be completely switched from waveguide 1 to waveguide 2 at the exit of the coupler [see the dashed curve in Fig. \ref{fig:pd}(a)]. However, if we tune the gate voltage to $V_{0}=129.86$ meV, the wave function initially injected into waveguide 1 will be equally split into the two waveguides at the exit of the coupler [see the dash-dotted curve in Fig. \ref{fig:pd}(c)]. Therefore, a voltage-controlled coherent device for Dirac electron switching and splitting can be implemented based on our scheme, and the size of the device is quite smaller than the observed coherence length ($\sim 1$ $\mu$m) in graphene. \cite{lcl}

\begin{table}
\caption{Numerical verification of the weak-coupling approximation. The parameters $R_{1}$, $R_{2}$, $R_{3}$, and $R_{4}$ for the weak-coupling approximation given by Eqs. (\ref{weak}) are numerically calculated, where the relationships of $R_{1}=R_{3}$ and $R_{2}=R_{4}$ result from the symmetric configuration of the coupling structure.}
\begin{tabular}{|r@{.}l|r@{.}l|r@{.}l|}
\hline \multicolumn{2}{|c|}{$\rule{0pt}{4ex} \ \ \ V_{0}$ (\text{meV})$\ \ \ $}   &   \multicolumn{2}{c|}{$\ \ \ R_{1}=R_{3} \ \ \ $} & \multicolumn{2}{c|}{$\ \ \ R_{2}=R_{4} \ \ \ $} \\ \hline
$\rule{0pt}{4ex} \ \ \ \ \ 139$&10          & $\ \ \ \ 0$&00830        & $\ \ \ \ \ 0$&104 \\ \hline
$\rule{0pt}{4ex}\ \ \ \ \ 129$&86          & $\ \ \ \ 0$&00340        & $\ \ \ \ \ 0$&052 \\ \hline
$\rule{0pt}{4ex} \ \ \ \ \ 121$&25          & $\ \ \ \ 0$&00074        & $\ \ \ \ \ 0$&031 \\ \hline
$\rule{0pt}{4ex} \ \ \ \ \ 31$&69           & $\ \ \ \ 0$&05300        & $\ \ \ \ \ 0$&041 \\ \hline
\end{tabular}
\label{table:r}
\end{table}

Finally, we should verify the validity of the weak-coupling approximation [i.e., Eqs. (\ref{weak})] in our scheme. Based on the derivations in Section II and the above geometric parameters of $w=45$ nm and $t=20$ nm, we can numerically calculate Eqs. (\ref{weak}) with different potential barriers $V_{0}$ and summarize the results in Table \ref{table:r}. One can clearly see that all the values of $R_{1}$, $R_{2}$, $R_{3}$, and $R_{4}$ are much smaller than unity and thus the weak-coupling approximation can be well satisfied with our proposed parameters. As a consequence, our investigations clearly indicate that the coupled-mode theory can effectively describe the coherent transport of Dirac electrons in waveguide-based directional coupling structures and can optimize the design of the quantum-field-effect-based coupling devices in graphene.

\section{Conclusion}
In conclusion, we have shown that a directional coupler consisting of two parallel waveguides can be constructed using local electrostatic gates in bulk graphene. In this coupler, by tuning a gate voltage, the wave function of Dirac electron initially injected into one waveguide can be partially or completely transferred to the other waveguide, which could lead to a voltage-controlled device for coherent electric current switching and splitting in graphene. More importantly, Klein tunneling can dramatically enhance the coupling strength between the waveguides, thereby shrinking the geometrical size of the coupler to be much smaller than the coherence length in graphene. Additionally, in analogy to the directional coupler in guided-wave optics, \cite{yariv} if a small potential difference is introduced between the two waveguides, multiplexing and demultiplexing devices for Dirac electron wave could also be implemented. Therefore, based on the Klein-tunneling-enhanced directional coupling structure, a variety of graphene-based compact logic devices could be realized for nanoelectronic applications in quantum-based electronic signal processing and on-chip device integration.

L.Z. thanks P. Tang and F. Peng for helpful discussions. We acknowledge the financial support from the Ministry of Science and Technology of China (Grants No. 2011CB921901 and 2011CB606405) and the National Science Foundation of China.

\bibliography{gdcarxiv}

\end{document}